\documentclass[aps,prd,twocolumn,preprintnumbers,superscriptaddress,nofootinbib]{revtex4-1}
\usepackage{graphicx}
\usepackage{epstopdf}
\usepackage{amsmath}
\usepackage{amsfonts}
\usepackage{amssymb}
\usepackage{appendix}
\usepackage{enumerate}
\usepackage{natbib}
\usepackage{comment}
\usepackage{bbold}
\usepackage[shortlabels]{enumitem}
\usepackage{color}
\usepackage{slashed}
\usepackage{subfigure}
\usepackage{setspace}
\usepackage{footnote}
\usepackage{lipsum}
\usepackage{multirow}
\usepackage[colorlinks = true,
            linkcolor = blue,
            urlcolor  = blue,
            citecolor = blue,
            anchorcolor = blue]{hyperref}
\usepackage[capitalize]{cleveref}
\usepackage{braket}
\usepackage{physics}
\usepackage[normalem]{ulem}
\usepackage{url}

\newcommand{\be}{\begin{eqnarray}}
\newcommand{\ee}{\end{eqnarray}}
\newcommand{\ba} {\begin{equation}\begin{aligned}}
\newcommand{\ea} {\end{aligned}\end{equation}}
\newcommand{\bg} {\begin{equation}\begin{gathered}}
\newcommand{\eg} {\end{gathered}\end{equation}}

\newcommand{\beq}{\begin{equation}}
\newcommand{\eeq}{\end{equation}}

\usepackage{orcidlink}
\begin{document}

\title{Solar Constraints on Heavy Neutral Leptons with $\nu_\tau$ Mixing}

\author{Vedran~Brdar\orcidlink{0000-0001-7027-5104}}
\email{vedran.brdar@okstate.edu}
\affiliation{Department of Physics, Oklahoma State University, Stillwater, OK 74078, USA}
\author{Samiur~R.~Mir\orcidlink{0000-0002-6531-2174}}
\email{samiur.mir@okstate.edu}
\affiliation{Department of Physics, Oklahoma State University, Stillwater, OK 74078, USA}
\author{Xun-Jie Xu\orcidlink{0000-0003-3181-1386}}
\email{xuxj@ihep.ac.cn}
\affiliation{Institute of High Energy Physics, Chinese Academy of Sciences, Beijing 100049, China}

\begin{abstract}
The mixing of heavy neutral leptons (HNLs) with tau neutrinos remains largely unconstrained compared to their mixing with electron and muon neutrinos. In this work, we investigate the potential of solar neutrinos to improve constraints on HNL-$\nu_\tau$ mixing in the MeV mass range. Due to neutrino oscillations, the Sun is a copious source of neutrinos of all flavors, and the partial conversion of electron neutrinos into muon and tau neutrinos occurs already during their propagation through the solar interior. Tau neutrinos can produce HNLs through scattering with protons in the Sun, provided nonzero HNL–$\nu_\tau$ mixing is present. The HNLs escape the Sun and subsequently decay through the same interaction. For $\mathcal{O}(10)$ MeV HNL masses, the dominant visible decay yields electrons and positrons that can be detected by space-based solar observatories. Using data from the Solar and Heliospheric Observatory (SOHO), we search for such signals and derive constraints on the squared HNL-$\nu_\tau$ mixing matrix element, $|U_{\tau N}|^2$, reaching below the $10^{-2}$ level for HNL masses of $\sim 5$ MeV. These constraints improve upon existing terrestrial limits in this mass range by more than an order of magnitude.
\end{abstract}

\maketitle
\textbf{Introduction.} 
\noindent
Heavy neutral leptons (HNLs), being gauge-singlet fermions, are among the simplest extensions of the Standard Model (SM).
Their interactions with the SM can arise through mixing with active neutrinos, resulting in a rich phenomenology accessible to a broad range of experimental probes~\cite{Atre:2009rg,deGouvea:2015euy,Bolton:2019pcu,Fernandez-Martinez:2023phj}. The mixing between an HNL ($N$) and an active neutrino $\nu_\alpha$ ($\alpha\in\{e,\mu,\tau\}$) is parametrized by the matrix element $U_{\alpha N}$ of the generalized lepton mixing matrix. The magnitude of the mixing controls the production and decay rates of HNLs. 

Currently, $|U_{eN}|^2$ and $|U_{\mu N}|^2$ are constrained to be $\lesssim \mathcal{O}(10^{-6})$ for HNL masses of $m_N\sim\mathcal{O}(10)~\mathrm{MeV}$~\cite{Borexino:2013bot,Drewes:2024dem,Daum:1987bg,PIENU:2019usb}. In contrast, to date, the most stringent laboratory constraints on $|U_{\tau 4}|^2$ are at the $\mathcal{O}(10^{-1})$ level in the same HNL mass range~\cite{Dentler:2018sju,Plestid:2020ssy,Gustafson:2022rsz}.
Therefore, among the three neutrino flavors, HNLs with $\nu_\tau$ mixing remain relatively unconstrained. The disparity can be attributed to the difficulty of producing $\tau$-flavored HNLs, as their production is generally associated with the production of $\nu_\tau$, which is often absent in terrestrial neutrino sources such as nuclear beta decays and meson decays.

Motivated by the lack of constraints in the $\nu_\tau$ sector, in this work, we consider MeV-scale Majorana HNLs produced in the Sun. Specifically, we show that
through \textit{neutrino oscillation and scattering}, the Sun can produce a sufficiently high flux of $\tau$-flavored HNLs. Subsequently, the HNLs decay in the heliosphere into final states containing electrons and positrons. We employ the data collected by the Solar and Heliospheric Observatory (SOHO)~\cite{SOHO1995S} to place constraints on HNLs with $\nu_\tau$ mixing.\footnote{Previously, heliospheric electron data were used to constrain HNL mixing with $\nu_e$, assuming HNL production from solar neutrinos through mixing~\cite{Drewes:2024dem}.} As we will show, our limits improve upon existing terrestrial constraints by more than an order of magnitude, making solar production the leading probe of HNL-$\nu_\tau$ mixing in the low-mass regime.

\begin{figure}[t!]
    \centering
       \includegraphics[width=0.95\linewidth]{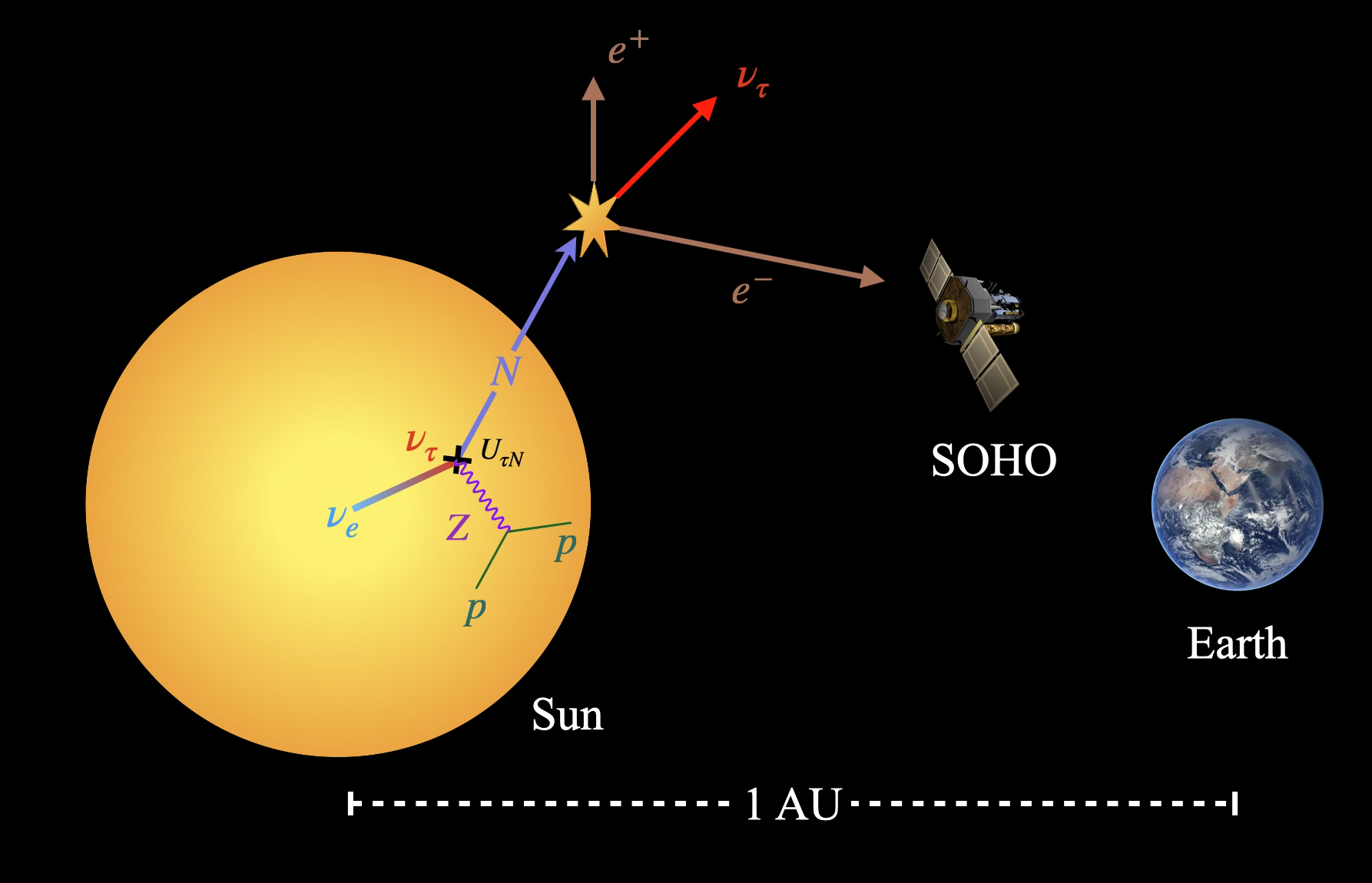}
    \caption{Illustration of production and decay of HNLs with $\nu_\tau$ mixing in the Sun. Electron neutrinos convert to $\nu_\tau$ via neutrino oscillations, followed by the production of HNLs through $\nu_\tau$ upscattering off nuclei. The produced HNLs escape the Sun and decay in the heliosphere into $e^+ e^- \nu_\tau$. The $e^\pm$ may be captured by space-based solar observatories (e.g., SOHO).  Earth and SOHO images are adapted from NASA public-domain imagery~\cite{NASA_Earth, SOHO_illustration}. } 
    \label{fig:illustration}
\end{figure}

\textbf{Solar HNL Production.}
\noindent
The solar production of $\tau$-flavored HNLs relies on a two-step mechanism: neutrino oscillation followed by neutrino scattering
with solar medium, as illustrated in \cref{fig:illustration}.
Neutrino oscillation alone cannot produce a heavy mass eigenstate
if that state is not initially produced at the source. However, oscillation
converts $\nu_{e}$ to $\nu_{\tau}$ during propagation, and the resulting
$\nu_{\tau}$ flux can subsequently produce HNLs through scattering
processes such as $\nu_{\tau}+p\to N+p$ in which the momentum
transfer is inevitably large.\footnote{The energy of the produced $N$ is approximately equal to that
of $\nu_{\tau}$, but its momentum, $p_{N}\approx(E_{\nu}^{2}-m_{N}^{2})^{1/2}$,
is significantly smaller than the momentum of $\nu_{\tau}$. Consequently,
the momentum transfer is always large, thereby precluding coherent
forward scattering. Hence, this large-momentum-transfer process cannot generate an MSW potential to interfere with neutrino oscillation.} A crucial difference between the solar production of $\tau$-flavored $N$
and $e$-flavored $N$ is that the latter can be readily
produced at the source via its mixing with $\nu_{e}$. From the perspective of mass
eigenstates, this difference lies in the absence or presence of a heavy
mass eigenstate at production. Oscillation only causes phase shifts
on mass eigenstates that have already been produced, without creating
new mass eigenstates or destroying existing ones. Therefore, without
scattering processes,  $\tau$-flavored $N$ cannot be produced via oscillation.

With the two-step mechanism, the production of $N$ is formulated
as follows
\begin{equation}
\frac{d\Phi_{N}^{(0)}}{dE_{N}}=\sum_{X}\int\frac{d\Phi_{\nu_{e}}}{dE_{\nu}}P_{e\tau}n_{X}\frac{d\sigma_{X}}{dE_{N}}dE_{\nu}dr\thinspace,\label{eq:HNL-prod}
\end{equation}
where $d\Phi_{N}^{(0)}/dE_{N}$ and $d\Phi_{\nu_{e}}/dE_{\nu}$ denote
the differential fluxes (i.e., the number of particles emitted per unit time and energy) of $N$ and $\nu_{e}$ with $E_{N}$
and $E_{\nu}$ being the respective energies; $P_{e\tau}$ is the
flavor transition probability of $\nu_{e}\to\nu_{\tau}$; $n_{X}$
is the number density of the target particle $X$, which can be a
nucleus or an electron; and $d\sigma_{X}/dE_{N}$ is the differential
cross section of $\nu_{\tau}+X\to N+X$. For such scattering in solar medium, the dominant contribution comes from $X=p$. The
superscript ``$(0)$'' in $d\Phi_{N}^{(0)}/dE_{N}$ indicates that no HNL decay effects are included, in contrast to $d\Phi_{N}/dE_{N}$ used in the next section. Assuming $X$ is sufficiently heavy, $N$
and $\nu_{\tau}$ in the scattering process should have approximately
the same energy, denoted by $E$. Under this approximation, \cref{eq:HNL-prod}
can be reduced to
\begin{equation}
\frac{d\Phi_{N}^{(0)}}{dE}\approx 4.1\times10^{-6}\,|U_{\tau N}|^2\frac{d\Phi_{\nu_{e}}}{dE}F_{S}(E,m_{N})\thinspace,\label{eq:-34}
\end{equation}
where 
$F_{S}(E,m_{N})\approx (E/15 {\rm MeV})^2 (1-m_N^2/E^2)^{1/2}$ accounts for spectral distortions. 
The numerical coefficient in \cref{eq:-34} is obtained by evaluating the flavor transition probability using the adiabatic approximation \cite{Haxton:1986dm,Parke:1986jy,Petcov:1987zj,deHolanda:2004fd} and performing the integration in \cref{eq:HNL-prod} using the standard solar profile \cite{Vinyoles:2016djt}; see the Supplemental Material~\ref{sec:HNLSupp}
for details.

\begin{figure}[t]
    \centering
       \includegraphics[width=0.95\linewidth]{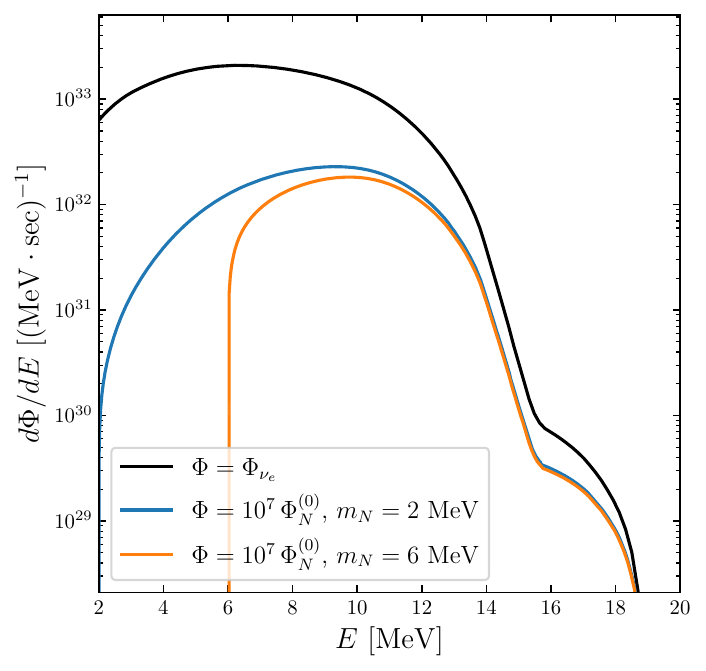}
        \caption{Solar HNL flux obtained using \cref{eq:-34} with $|U_{\tau N}|^2=10^{-2}$ and $m_N\in \{2, 6\}\,{\rm MeV}$ (blue and orange lines), to be compared with the solar $\nu_e$ flux (black line). Note that the solar HNL flux has been multiplied by a factor of $10^{7}$ for better visualization and it does not include the decay effect.
        } 
    \label{fig:flux}
\end{figure}

In \cref{fig:flux}, we present the solar $\tau$-flavored HNL flux obtained using \cref{eq:-34} for $m_N\in \{2, 6\}\,{\rm MeV}$ and $|U_{\tau N}|^2=10^{-2}$. For comparison, the original solar $\nu_e$ flux (without oscillation) is also presented. Here we only include the contributions of $^{8}{\rm B}$ and hep neutrinos. At energies below the range shown in the figure, pp and pep neutrinos could make prominent contributions to the production of sub-MeV HNLs. However, these neutrinos do not contribute to our analysis, which requires $N$ to be heavier than twice the electron mass to produce the observable $e^\pm$ signal. 

\textbf{HNL Decay.} 
\noindent
Given their relatively long lifetimes for $\mathcal{O}(10)$ MeV masses, HNLs produced in the Sun typically decay in the heliosphere. Taking decay into account, the HNL flux at a heliocentric distance $r$ reads
\begin{equation}
\frac{d\Phi_{N}}{dE_{N}}=\frac{d\Phi_{N}^{(0)}}{dE_{N}}e^{-r/R_{D}}\thinspace,\label{eq:decay}
\end{equation}
where $R_{D}=\tau_{N}v_{N} (1-v_{N}^{2})^{-1/2}$ is the mean
flight distance, with $\tau_{N}$ the decay-at-rest lifetime 
and $v_{N}$ the velocity of $N$. 

The produced HNLs must be lighter than the maximum energy of the solar neutrino spectrum, $18.77$ MeV, which is below the masses of any mesons or heavy charged leptons ($\mu,\tau$). 
Hence, HNLs can only decay via three channels, $N \to \nu_\tau \gamma$, $N \to \nu_\alpha \bar{\nu}_\alpha \nu_\tau$ and $N \to e^+ e^- \nu_\tau$, with the last two being the dominant ones. The relevant decay widths in the HNL rest frame are given in Ref.~\cite{Gorbunov:2007ak}. We are interested in the $N \to e^+ e^- \nu_\tau$ decay channel because of the $e^\pm$ detection prospects. Namely, this decay channel may 
result in an excess of $e^\pm$ flux in the heliosphere. Previously, such signal was used to constrain HNL mixing with electron neutrinos~\cite{Drewes:2024dem}. In what follows, we discuss the $e^\pm$ spectrum induced by $\tau$-flavored HNLs and derive constraints on HNL-$\nu_\tau$ mixing.

The propagation of $e^\pm$ produced from solar HNL decay is strongly influenced by the magnetic field in the heliosphere. 
Their typical mean free path within the relevant energy range is well below 1~AU, implying that the $e^{\pm}$ flux to be observed near Earth is a consequence of diffusion. Neglecting adiabatic energy loss and the solar wind effect, the diffusion process is described by the following transport equation~\cite{Engelbrecht2022S}
\begin{equation}
    \frac{\partial \psi}{\partial t} = \nabla \cdot (\textbf{D}\cdot\nabla \psi) + q \,,
    \label{eq:psi_diffusion}
\end{equation}
where $\psi=dn_e/dE_e$ denotes the number density of $e^{\pm}$ per unit energy. On the RHS, the first term represents the diffusion, accounting for particle transport from high- to low-concentration locations, weighted by the diffusion tensor $\textbf{D}$. 
The source term, $q$, gives the number of electrons and positrons of energy $E_e$ produced by solar HNL decay per unit volume, time, and energy at a heliocentric distance $r$. It is computed as follows \cite{Drewes:2024dem}
\begin{equation}
q(r,E_e) = \frac{2}{4\pi r^{2}}\int dE_{N}\frac{1}{v_{N}}\frac{d\Phi_{N}}{dE_{N}}\frac{d\Gamma^{\rm flight}_{N \to e^+ e^- \nu_\tau }}{dE_{e}}\,,
\label{eq:src_q}
\end{equation} 
where ${d\Gamma^{\rm flight}_{N \to e^+ e^- \nu_\tau }}/{dE_e}$ is the differential decay rate of boosted HNL with energy $E_N$. The factor of 2 reflects the inclusion of both $e^+$ and $e^-$ in the source term.  
We note here that for a given value of $E_e$, the kinematically allowed range of $E_N$ not only has a lower bound but also has an upper bound due to the finite mass of the electron; for more details, see the Supplemental Material~\ref{sec:decay-kinematics}.

Let us briefly comment on the calculation of ${d\Gamma^{\rm flight}_{N \to e^+ e^- \nu_\tau }}/{dE_{e}}$.
A generic HNL mixed with all neutrino flavors can decay into $e^+  e^-\nu_\alpha$ with $\alpha\in \{e,\mu,\tau\}$. 
For $\alpha=e$, both charged current (CC) and neutral current (NC) interactions are involved, while for  $\alpha\neq e$, the decay only goes through NC interaction. 
In the former case, the left-chiral coupling reads $g_L={1}/{2}+s_W^2$, while the latter case corresponds to $g_L=-{1}/{2}+s_W^2$ ($s_W\equiv \sin \theta_W$ where $\theta_W$ is the weak mixing angle). 
Because of the sign flip in front of the factor of $1/2$ in $g_L$, the difference between the CC+NC and NC-only cases becomes apparent at the amplitude-squared level containing $s_W^2$ and $s_W^4$; the ratio of the coefficient of $s_W^2$ to that of $s_W^4$ is $1/2$ ($-1/2$) for $\alpha = e$ ($\alpha \neq e$), as given in Table~3 of Ref.~\cite{Bondarenko:2018ptm}. 
This allows us to adapt the differential decay rate of $N \to e^+ e^- \nu_e$ previously obtained in Ref.~\cite{Drewes:2024dem} into a formula applicable to $N \to e^+ e^- \nu_\tau$ by  mapping $U_{eN} \to U_{\tau N}$ and $s_W^2 + 2 s_W^4 \to -s_W^2 + 2 s_W^4$.

Over a sufficiently long time scale, the diffusion process develops a steady-state profile of $\psi$, implying ${\partial \psi}/{\partial t}=0$. Assuming that the diffusion is spherically symmetric, the diffusion term  $\nabla \cdot (\textbf{D}\cdot\nabla \psi)$ in \cref{eq:psi_diffusion} simplifies to $D \nabla^2\psi$ where $D\approx 2\times 10^{22}\,\rm cm^2 \, s^{-1}$ is the diffusion coefficient~\cite{Drewes:2024dem}. 
\cref{eq:psi_diffusion} can then be solved using the standard Green's function method, resulting in
\begin{equation}
    \psi(E_e) = \frac{\xi(E_e)}{D\, r} \left[e^{-\frac{R_\odot}{R_D}} - e^{-\frac{r}{R_D}} + \frac{r}{R_D} \, \Gamma\left(0, \frac{r}{R_D}\right) \right]\,,
    \label{eq:psi_comp}
\end{equation}
where $\Gamma$ is the incomplete Gamma function. 
In deriving this equation, we include only the $r>R_\odot$ contribution, as HNLs decaying within the solar medium cannot contribute to the heliospheric $e^{\pm}$ flux. 
The $\xi(E_e)$ function denotes the energy dependence of $q$ which is approximately factorized as $q\approx \xi(E_e)e^{-r/R_D}/(R_D r^2)$~\cite{Drewes:2024dem}.

The $e^{\pm}$ signal observed by, e.g., space-based solar observatories is characterized by the differential intensity, $J_{\rm obs}$, defined as the number of particles observed per unit time, per unit area, per unit energy, and within a unit solid angle. For the diffusing $e^{\pm}$ flux, which becomes isotropic at the observation site, $J_{\rm obs}$ is related to $\psi$ by
\begin{equation}
    J_{\rm obs}  = \frac{v_e}{4\pi} \psi\,,
    \label{eq:Jobs}
\end{equation}
where $v_e$ is the electron velocity.

\begin{figure}[t!]
    \centering
       \includegraphics[width=0.95\linewidth]{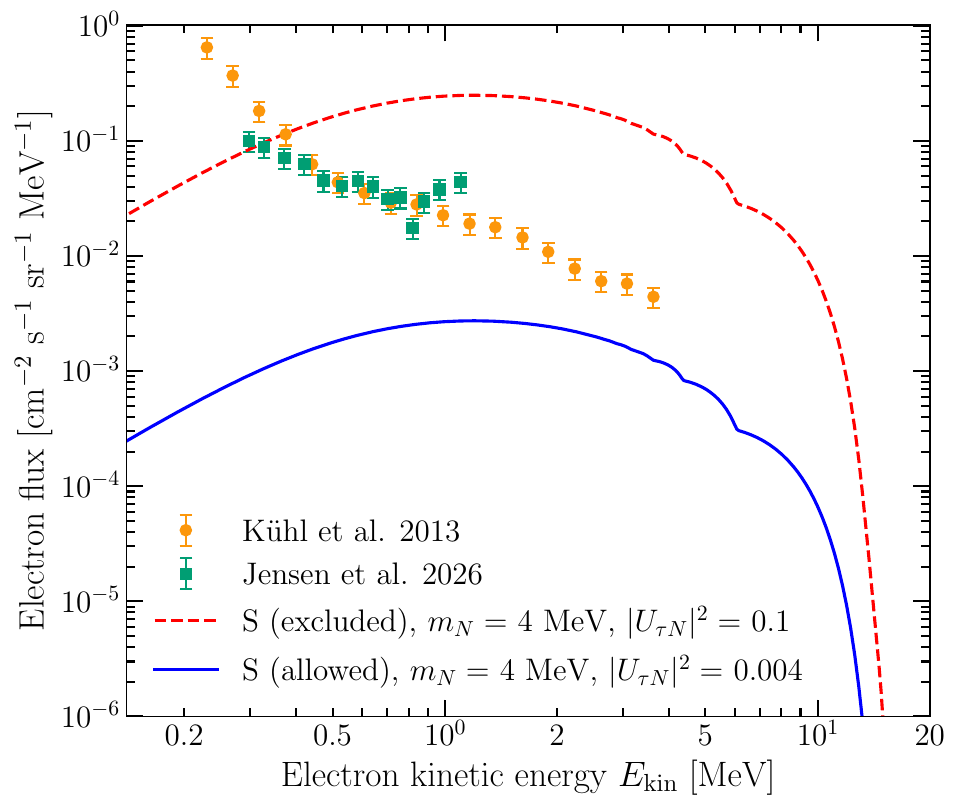}
    \caption{Electron fluxes from $N$ decays for two benchmark points are compared with the SOHO data~\cite{Kuhl:2013icrc,Jensen2026SOHO}. The red dashed and blue solid curves correspond to excluded and allowed values of $|U_{\tau N}|^2$, respectively. These benchmark points are also indicated in \cref{fig:constraint} by a red star and a blue diamond.} 
    \label{fig:signal_data}
\end{figure}

\textbf{Results.}
\noindent
In our analysis, we use data collected by the space-based probe SOHO~\cite{SOHO1995S}. SOHO operates in an orbit near Earth, at a distance of approximately 1~AU from the Sun. The Electron, Proton, Helium Instrument (EPHIN) \cite{COSTEP} onboard SOHO measures the combined electron and positron spectrum without distinguishing between the two species.

Energetic charged particles from the Sun associated with solar activity can leave a similar signature at SOHO as $e^+e^-$ pairs from HNL decays. However, solar activity declines over each solar cycle, reaching a period known as the ``Quiet Sun.'' The reduced solar-wind background during this period makes it particularly well suited for searches for heliospheric electrons and positrons from solar HNL decays.

SOHO collected MeV-electron data (hereafter, ``electron'' refers to both $e^-$ and $e^+$) over a 13-month period from June 2007 to July 2008~\cite{Kuhl:2013icrc,Jensen2026SOHO}, during the declining phase of Solar Cycle 23 and leading into the deep solar minimum between Solar Cycles 23 and 24~\cite{NOAA:solarcycle}. This measurement, employed in our analysis, is dominated by Jovian electrons from Jupiter and galactic electrons.

In \cref{fig:signal_data}, we compare the SOHO data with the electron signal from $\tau$-flavored HNL decays computed using \cref{eq:psi_comp,eq:Jobs} for two benchmark points. The two datasets from Ref.~\cite{Kuhl:2013icrc} and Ref.~\cite{Jensen2026SOHO} are shown as orange and green data points, respectively. While both datasets correspond to the same SOHO measurement, the latter incorporates an improved treatment of the electron response function. In our analysis, we employ the data from Ref.~\cite{Kuhl:2013icrc}, which cover a substantially broader range of electron kinetic energies. Regarding the signal, the representative benchmark points are shown by the blue solid ($|U_{\tau N}|^2=0.004$) and red dashed ($|U_{\tau N}|^2=0.1$) curves, respectively. They are chosen such that the former is clearly allowed, while the latter is excluded by the data, as the electron flux from HNL decays exceeds the observed flux over most of the energy range.

\begin{figure}[t!]
    \centering
       \includegraphics[width=\linewidth]{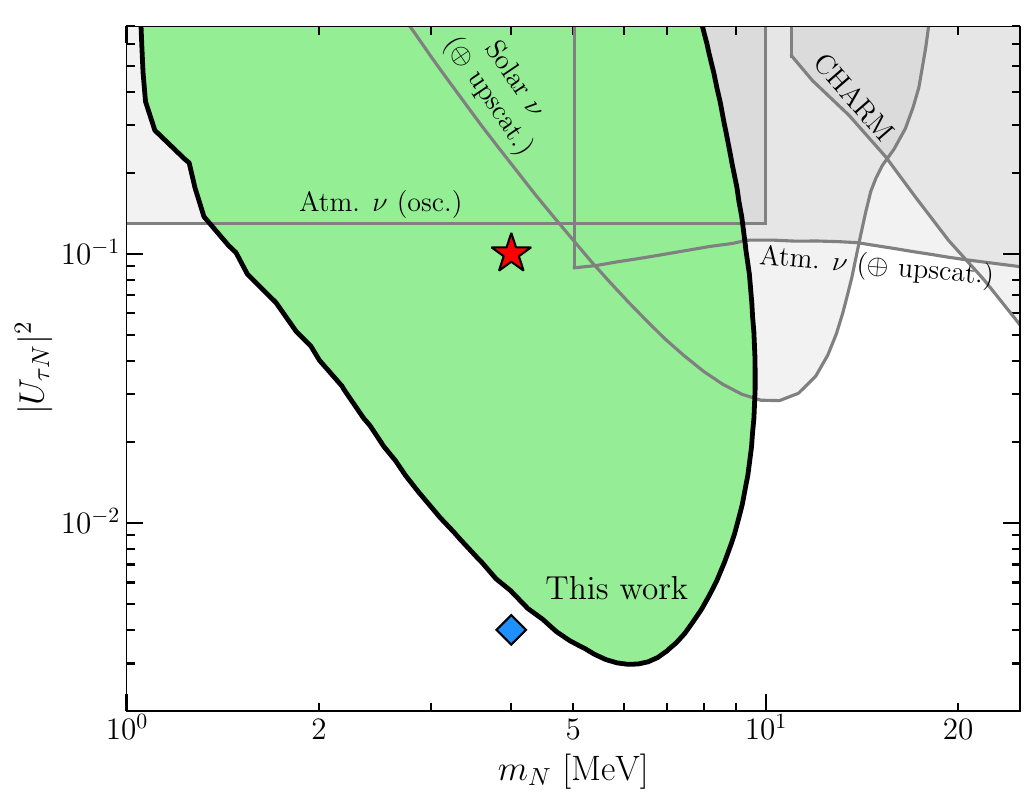}
    \caption{Our constraints at 90\% CL in the parameter space of the HNL mass and the HNL mixing with tau neutrino are shown in green. We compare our results with other relevant constraints obtained from solar neutrinos (earth upscattering~\cite{Plestid:2020ssy}),  atmospheric neutrinos (oscillation~\cite{Dentler:2018sju} and earth upscattering~\cite{Gustafson:2022rsz}), and CHARM \cite{Orloff:2002de}. Two benchmark points, for which we present the $e^\pm$ flux in \cref{fig:signal_data}, are highlighted by a red star and a blue diamond.} 
    \label{fig:constraint}
\end{figure}
The above benchmark points provide an illustrative comparison between the HNL signal and the SOHO data. To derive constraints in the $m_N$-$|U_{\tau N}|^2$ parameter space, we employ the following test statistic
\begin{equation}
    \Delta \chi^2 = \sum_i \frac{(S_{i}  +B_{i}-O_{i})^2}{\sigma_{i}^2}\,,
    \label{eq:chi2}
\end{equation}
where $i$ labels the energy bins, $S_i$ and $B_i$ denote the expected signal and background in the $i$-th bin, respectively, while $O_i$ denotes the measured flux and $\sigma_i$ its uncertainty. We employ the data shown in orange in \cref{fig:signal_data}, and assume a 20\% relative uncertainty, i.e., $\sigma_i=0.2\,O_i$. This choice provides a conservative estimate of the experimental uncertainty~\cite{Jensen2026}. The signal $S_i$ depends on the HNL parameters and is computed using \cref{eq:psi_comp,eq:Jobs}. We conservatively assume that the observed flux can be entirely attributed to background processes. Thus, we set $B_i=O_i$ and interpret any additional contribution from HNL decays as a potential excess over the observed flux. This allows us to derive constraints in the $m_N$-$|U_{\tau N}|^2$ parameter space.

Our derived 90\% CL constraints are presented in \cref{fig:constraint}. For comparison, we also show in the figure the strongest existing terrestrial constraints on low-mass $\tau$-flavored HNLs, obtained from atmospheric and solar neutrino data~\cite{Dentler:2018sju, Plestid:2020ssy, Gustafson:2022rsz}, as well as from the CHARM experiment~\cite{Orloff:2002de}. 
In particular, the disappearance of atmospheric neutrinos studied in Ref.~\cite{Dentler:2018sju} yields an $m_N$-independent constraint of $|U_{\tau N}|^2 < 0.13$ at 90\% CL. Neutrinos can also upscatter within the Earth, producing $\tau$-flavored HNLs that propagate and decay inside a detector. The resulting constraints have been derived using solar neutrinos with Borexino~\cite{Plestid:2020ssy} and atmospheric neutrinos with Super-Kamiokande~\cite{Gustafson:2022rsz}. The CHARM experiment provides constraints reaching $|U_{\tau N}|^2\simeq 10^{-6}$ for $\mathcal{O}(1)$~GeV HNL masses~\cite{Boiarska:2021yho}, but these constraints become relatively weak below $m_N=20$~MeV as can be seen in \cref{fig:constraint}. Our results improve upon all of these existing limits by more than an order of magnitude in $|U_{\tau N}|^2$ at $m_N\sim 5$~MeV.

We note that Big Bang Nucleosynthesis (BBN) observations could put stringent constraints on HNL mixing~\cite{Ruchayskiy:2012si, Sabti:2020yrt, Boyarsky:2020dzc,Chen:2024cla,Dev:2025pru}. However, these limits depend on cosmological models and can be significantly weakened if the early universe has a low reheating temperature.  
Supernovae provide a complementary probe of HNL mixing. While traditional supernova cooling constraints do not reach our parameter space of interest, bounds from low-explosion-energy supernovae overlap with our sensitivity region~\cite{Carenza:2023old}. However, these bounds exhibit dependence on the assumed supernova envelope size, progenitor model, and treatment of HNL energy deposition~\cite{Chauhan:2023sci}.

\textbf{Summary and Conclusions.}
\noindent
In this work, we calculated, for the first time, the $\nu_\tau$ flux in the Sun's interior and considered their upscattering to heavy neutral leptons (HNLs) through mixing. The produced HNLs escape the Sun and subsequently decay through the same interaction. This mechanism probes $\tau$-flavored HNL masses up to $\mathcal{O}(10)~\mathrm{MeV}$ for which the dominant visible decay products are electrons and positrons. These charged particles can be detected by space-based solar observatories, and in this work, we employed data collected by SOHO to search for their signatures. The absence of an anomalous excess in the SOHO data allows us to derive new constraints on the $\tau$-flavored HNL parameter space, reaching below $|U_{\tau N}|^2\sim 10^{-2}$ at $m_N\sim 5$ MeV. This limit surpasses all previously derived terrestrial constraints for HNL masses in the $1$--$20~\mathrm{MeV}$ range. Our results establish solar neutrinos, combined with space-based observations, as a novel and robust probe of HNLs with dominant $\nu_\tau$ mixing, opening a new avenue for searches for physics beyond the Standard Model.

\textbf{Acknowledgments.} 
\noindent
SRM is grateful to Stefan Jensen for discussions regarding the SOHO/EPHIN data and to Bernd Heber for valuable comments. The work of VB and SRM is supported by the United States Department of Energy Grant No. DE-SC0016013.  XJX is supported in part by the National Natural Science Foundation of China under grant No.~12141501 and also by the CAS Project for Young Scientists in Basic Research (YSBR-099).

\bibliographystyle{JHEP}
\bibliography{refs}

\clearpage
\newpage
\onecolumngrid

\setcounter{equation}{0}
\setcounter{figure}{0}
\setcounter{table}{0}
\setcounter{page}{1}
\setcounter{section}{0}

\makeatletter
\renewcommand{\thesection}{S\arabic{section}}
\renewcommand{\theequation}{S\arabic{equation}}
\renewcommand{\thefigure}{S\arabic{figure}}
\renewcommand{\thetable}{S\arabic{table}}

\renewcommand{\theHfigure}{S\arabic{figure}}
\renewcommand{\theHtable}{S\arabic{table}}
\renewcommand{\theHequation}{S\arabic{equation}}
\makeatother

\begin{center}
    \textbf{\large Supplemental Material: Solar Constraints on Heavy Neutral Leptons with $\nu_\tau$ Mixing} \\ 
    \vspace{0.5cm}
    {Vedran~Brdar, Samiur~R.~Mir, Xun-Jie Xu}
\end{center}

\section{HNL production through oscillation and scattering \label{sec:HNLSupp}}

A rigorous treatment of the production of $\tau$-flavored HNLs through
the interplay of oscillation and scattering requires a careful consideration
of flavor and mass eigenstates, as well as the charged- and neutral-current
interactions involving these states. Without properly accounting for
these aspects, misconceptions regarding whether $N$ can be
directly produced through oscillation may arise. Below we present
a rigorous treatment of the production. 

First, we denote neutrino flavor and mass eigenstates by $(\nu_{e},\ \nu_{\mu},\ \nu_{\tau},\ N)$
and $(\nu_{1},\ \nu_{2},\ \nu_{3},\ \nu_{4})$, respectively. Here,
$(\nu_{e},\nu_{\mu},\nu_{\tau})$ have certain Standard Model (SM) charges and $N$
is a SM singlet. By contrast, the mass eigenstates $(\nu_{1},\ \nu_{2},\ \nu_{3},\ \nu_{4})$
do not have well-defined SM charges but they have well-defined masses.
 The two sets of states are related to each other by 
\begin{equation}
\left(\begin{array}{c}
\nu_{e}\\
\nu_{\mu}\\
\nu_{\tau}\\
N
\end{array}\right)=\left(\begin{array}{cccc}
1\\
 & 1\\
 &  & c_{\theta} & s_{\theta}\\
 &  & -s_{\theta} & c_{\theta}
\end{array}\right)\left(\begin{array}{cccc}
U_{e1}^{\text{PMNS}} & U_{e2}^{\text{PMNS}} & U_{e3}^{\text{PMNS}}\\
U_{\mu1}^{\text{PMNS}} & U_{\mu2}^{\text{PMNS}} & U_{\mu3}^{\text{PMNS}}\\
U_{\tau1}^{\text{PMNS}} & U_{\tau2}^{\text{PMNS}} & U_{\tau3}^{\text{PMNS}}\\
 &  &  & 1
\end{array}\right)\left(\begin{array}{c}
\nu_{1}\\
\nu_{2}\\
\nu_{3}\\
\nu_{4}
\end{array}\right),\label{eq:-28}
\end{equation}
where $U_{\alpha i}^{\text{PMNS}}$, with $\alpha\in\{e,\ \mu,\ \tau\}$
and $i\in\{1,\ 2,\ 3\}$, denotes the lepton mixing (PMNS) matrix, and $(c_{\theta},\ s_{\theta})\equiv(\cos\theta,\ \sin\theta)$,
with $\theta$ a small angle accounting for the $\tau$-flavored HNL
mixing. Denoting the product of the two matrices in \cref{eq:-28}
by $U$, we have $\nu_{\alpha}=\sum_{i=1}^{4}U_{\alpha i}\nu_{i}$
with 
\begin{equation}
U_{e4}=U_{\mu4}=0\thinspace,\ \ U_{\tau4}=s_{\theta}\thinspace.\label{eq:-29}
\end{equation}
\cref{eq:-29} implies that $\nu_{e}$ and $\nu_{\mu}$ do not
contain the heaviest mass eigenstate $\nu_{4}$ at all. Therefore,
in the case of $\tau$-flavored HNL mixing, $\nu_{4}$ cannot be produced
by any nuclear reactions in the Sun,  even if its mass is sufficiently
light and kinematically accessible. 

In contrast to $\nu_{e}$ and $\nu_{\mu}$, $\nu_{\tau}$ contains
a small fraction of $\nu_{4}$. As a consequence, when oscillation
converts a neutrino from the electron flavor to the tau flavor, the
conversion is, strictly speaking, not $\nu_{e}\to\nu_{\tau}$ because
$\nu_{\tau}$ contains a heavy component. Instead, we have $\nu_{e}\to\nu'_{\tau}$
with $\nu'_{\tau}$ defined by
\begin{equation}
\nu'_{\tau}\equiv\sum_{i=1}^{3}U_{\tau i}^{\text{PMNS}}\nu_{i}\thinspace.\label{eq:-30}
\end{equation}
With nonzero $s_{\theta}$, obviously
\begin{equation}
\nu'_{\tau}\neq\nu_{\tau}\thinspace.\label{eq:-31}
\end{equation}

Note that $(\nu_{e},\ \nu_{\mu},\ \nu'_{\tau},\ \nu_{4})$ forms an
orthogonal and complete basis. Hence the standard framework of solar
neutrino oscillation can still be used---one only needs to replace
$\nu_{\tau}$ by $\nu'_{\tau}$ and omit $\nu_{4}$ because it cannot
be produced at the source and therefore remains absent during oscillation. 

With only $\nu_{1}$, $\nu_{2}$ and $\nu_{3}$ involved in neutrino
oscillation, the conversion probability of $\nu_{e}\to\nu'_{\tau}$
under the adiabatic approximation reads~\cite{Xu:2022wcq}
\begin{equation}
P_{e\tau}(r)=\sum_{i=1}^{3}|U_{ei}^{(m)}(0)|^{2}|U_{\tau i}^{(m)}(r)|^{2},\label{eq:-32}
\end{equation}
where $U^{(m)}$ denotes the effective mixing matrix in solar medium
which is $r$-dependent. In turn, this leads to the $r$-dependence of $P_{e\tau}$. 

Since oscillation produces $\nu'_{\tau}$ instead of $\nu_{\tau}$,
we need to reformulate the neutral-current interactions of neutrinos
in the $(\nu_{e},\ \nu_{\mu},\ \nu'_{\tau},\ \nu_{4})$ basis.
This can be obtained through the following basis transformation
\begin{equation}
\left(\begin{array}{c}
\nu_{e}\\
\nu_{\mu}\\
\nu_{\tau}\\
N
\end{array}\right)^{\dagger}\left(\begin{array}{cccc}
1\\
 & 1\\
 &  & 1\\
 &  &  & 0
\end{array}\right)\left(\begin{array}{c}
\nu_{e}\\
\nu_{\mu}\\
\nu_{\tau}\\
N
\end{array}\right)=\left(\begin{array}{c}
\nu_{e}\\
\nu_{\mu}\\
\nu'_{\tau}\\
\nu_{4}
\end{array}\right)^{\dagger}\left(\begin{array}{cccc}
1\\
 & 1\\
 &  & c_{\theta}^{2} & c_{\theta}s_{\theta}\\
 &  & c_{\theta}s_{\theta} & s_{\theta}^{2}
\end{array}\right)\left(\begin{array}{c}
\nu_{e}\\
\nu_{\mu}\\
\nu'_{\tau}\\
\nu_{4}
\end{array}\right),\label{eq:-33}
\end{equation}
which implies that the neutral-current interactions of $\nu_{e}$
and $\nu_{\mu}$ are not modified. For $\nu'_{\tau}$ and $\nu_{4}$,
the neutral-current interactions become non-diagonal, thereby allowing
for the production of $\nu_{4}$ via the process $\nu'_{\tau}+p\to\nu{}_{4}+p$. 

In a similar way, one can also show that the charged-current interactions
of $(\nu_{e},\ e)$ and $(\nu_{\mu},\ \mu)$ are not modified. Hence,
neutrino production in the Sun through nuclear reactions
is not modified. 

From \cref{eq:-33}, we observe that the cross section of $\nu'_{\tau}+p\to\nu{}_{4}+p$
can be readily obtained from the cross section of the SM process,
$\nu{}_{\tau}+p\to\nu_{\tau}+p$, which has been calculated in Ref.~\cite{Beacom:2002hs}.
The two cross sections differ by a factor of $s_{\theta}^{2}\approx\theta^{2}$
and a phase space suppression factor $\sqrt{1-m_{N}^{2}/E_{N}^{2}}$. 

Substituting the conversion probability and the cross section into
\cref{eq:HNL-prod} and performing the integration using the
standard solar profile from Ref.~\cite{Vinyoles:2016djt}, we obtain
the result shown in \cref{eq:-34}.

\section{HNL decay kinematics \label{sec:decay-kinematics}}

The three-body decay process $N\to e^{+}e^{-}\nu_{\tau}$ involves
relatively complex kinematics when $N$ is boosted. In particular,
it exhibits a counterintuitive feature: for a given value of the electron
energy $E_{e}$, there is an upper bound on $E_{N}$ to generate such
an electron in the kinematically allowed range. This feature arises
from the finite electron mass and should vanish in the limit of $m_{e}\to0$.
Below we present a detailed analysis of this kinematics. 

Without loss of generality, we assume that the momenta of $N$ and
$e^{-}$ can be written as 
\begin{align}
p_{N}^{\mu} & =(E_{N},\ 0,\ 0,\ p_{N})\thinspace,\label{eq:-35}\\
p_{e}^{\mu} & =(E_{e},\ 0,\ p_{e}s_{\theta},\ p_{e}c_{\theta})\thinspace,\label{eq:-36}
\end{align}
where $(s_{\theta},c_{\theta})\equiv(\sin\theta,\ \cos\theta)$, with $\theta$ denoting the angle between the outgoing $e^{-}$ and the boosted $N$.
 We denote the momenta of the other two final states by $p_{2}^{\mu}$
and $p_{3}^{\mu}$, which do not necessarily lie in the $y$-$z$
plane. But their sum, $q^{\mu}\equiv p_{2}^{\mu}+p_{3}^{\mu}$, must
lie in the $y$-$z$ plane. Furthermore, it is straightforward to
prove that the invariant mass of the two final states has a minimum:
\begin{equation}
q^{2}=(p_{2}+p_{3})^{2}=2p_{2}\cdot p_{3}+m_{e}^{2}\geq m_{e}^{2}\thinspace.\label{eq:-37}
\end{equation}
The invariant mass reaches the minimum when the two final states are
collinear. 

\cref{eq:-37} imposes a crucial constraint on $p_{N}$, $p_{e}$,
and $\theta$.  To make this explicit, we substitute $q^{\mu}=p_{N}^{\mu}-p_{e}^{\mu}$
into \cref{eq:-37} and obtain
\begin{equation}
m_{N}^{2}+2p_{N}p_{e}c_{\theta}-2E_{N}E_{e}\geq0\thinspace,\label{eq:-38}
\end{equation}
which implies
\begin{equation}
c_{\theta}\geq c_{\theta}^{{\rm \min}}\ \ \text{with}\ \ c_{\theta}^{{\rm \min}}=\frac{2E_{N}E_{e}-m_{N}^{2}}{2p_{N}p_{e}}\thinspace.\label{eq:-39}
\end{equation}

Note that the second term in \cref{eq:-38} is always smaller
than the third term, implying that their sum $2p_{N}p_{e}\cos\theta-2E_{N}E_{e}<2(p_{N}\cos\theta-E_{N})p_{e}$
could become very large and negative if $p_{e}$ is too large. Hence,
large $p_{e}$ could potentially render the left-hand side of \cref{eq:-38}
negative, thereby breaking down the kinematic constraint.    The
maximally allowed value of $p_{e}$ can be obtained by solving $c_{\theta}^{{\rm \min}}=1$
with respect to $p_{e}$, resulting in 
\begin{equation}
p_{e}^{\max}=\frac{p_{N}}{2}+\frac{E_{N}}{2}\sqrt{1-\frac{4m_{e}^{2}}{m_{N}^{2}}}\thinspace.\label{eq:-40}
\end{equation}

For fixed $p_{N}$, $p_{e}$ should satisfy $p_{e}\leq p_{e}^{\max}$, otherwise $c_{\theta}^{{\rm \min}}$ would exceed $1$, leaving
no kinematically allowed interval for $c_{\theta}$. 

\cref{eq:-38} can also be used to derive an upper bound on $p_{N}$
when $p_{e}$ is fixed.  Following a similar analysis, we obtain
\begin{equation}
p_{N}^{\max}=\frac{m_{N}^{2}}{m_{e}^{2}}\left(\frac{p_{e}}{2}+\frac{E_{e}}{2}\sqrt{1-\frac{4m_{e}^{2}}{m_{N}^{2}}}\right)\thinspace.\label{eq:-40-1}
\end{equation}
This may seem counterintuitive, as it implies that a sufficiently
boosted HNL is unable to produce a low-energy electron. In fact, it
is an effect caused by the nonzero electron mass. In the limit of
$m_{e}\to0$, $p_{N}^{\max}$ in \cref{eq:-40-1} approaches
$\infty$ and therefore can be neglected. 

From \cref{eq:-40} and \cref{eq:-40-1}, it is straightforward
to obtain the corresponding upper bounds on $E_{e}$ (for fixed $E_{N}$)
and on $E_{N}$ (for fixed $E_{e}$), given by $E_{e}^{\max}=\sqrt{(p_{e}^{\max})^{2}+m_{e}^{2}}$
and $E_{N}^{\max}=\sqrt{(p_{N}^{\max})^{2}+m_{N}^{2}}$. 

\begin{figure}
\centering
\includegraphics[width=0.5\textwidth]{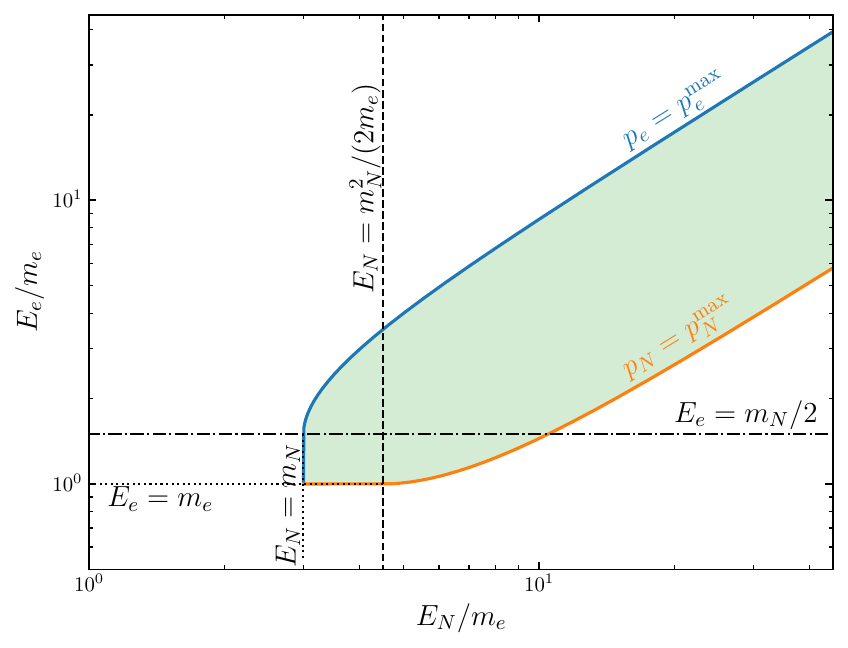}
\caption{The kinematically allowed ranges of $E_{e}$ and $E_{N}$. \label{fig:kinematics}}
\end{figure}

In \cref{fig:kinematics}, we present the corresponding upper
bounds on $E_{e}$ (for fixed $E_{N}$) and on $E_{N}$ (for fixed
$E_{e}$) in the $E_{e}$-$E_{N}$ plane, assuming $m_{N}=3m_{e}$
for illustration. For fixed $E_{N}$, $E_{e}$ should lie below the
blue curve. For fixed $E_{e}$, $E_{N}$ can only take values in the region to the left of the orange curve. The two requirements combined together indicate
the kinematically allowed region, which is presented in light green. 

From \cref{fig:kinematics}, one can see that the lower bound
on $E_{e}$ for fixed $E_{N}$ corresponds to the upper bound on $E_{N}$
for fixed $E_{e}$, and vice versa.  Therefore, for $E_{N}$ fixed
at a given value, the kinematically allowed range of $E_{e}$ is given
by $E_{e}^{\min}\leq E_{e}\leq E_{e}^{\max}$ with 
\begin{align}
E_{e}^{\min} & =\frac{1}{2}\left(E_{N}-p_{N}\sqrt{1-\frac{4m_{e}^{2}}{m_{N}^{2}}}\right)\Theta_{e}+m_{e}\left(1-\Theta_{e}\right),\label{eq:-22}\\
E_{e}^{\max} & =\frac{1}{2}\left(E_{N}+p_{N}\sqrt{1-\frac{4m_{e}^{2}}{m_{N}^{2}}}\right),\label{eq:-23}
\end{align}
where $\Theta_{e}=1$ or $0$ for $E_{N}-\frac{m_{N}^{2}}{2m_{e}}>0$
or $\leq 0$, respectively.

For fixed $E_{e}$, the kinematically allowed range
of $E_{N}$ is given by $E_{N}^{\min}\leq E_{N}\leq E_{N}^{\max}$
with 
\begin{align}
E_{N}^{\min} & =\frac{m_{N}^{2}}{2m_{e}^{2}}\left(E_{e}-p_{e}\sqrt{1-\frac{4m_{e}^{2}}{m_{N}^{2}}}\right)\Theta_{N}+m_{N}\left(1-\Theta_{N}\right),\label{eq:-22-1}\\
E_{N}^{\max} & =\frac{m_{N}^{2}}{2m_{e}^{2}}\left(E_{e}+p_{e}\sqrt{1-\frac{4m_{e}^{2}}{m_{N}^{2}}}\right),\label{eq:-23-1}
\end{align}
where $\Theta_{N}=1$ or $0$ for $E_{e}-m_{N}/2>0$ or $\leq 0$,
respectively. \\

In the calculation of the electron energy spectrum, we first integrate
over $\cos\theta$ 

\begin{equation}
\frac{d\Gamma_{N\to e^{+}e^{-}\nu_{\tau}}^{{\rm flight}}}{dE_{e}}=\int_{c_{\theta}^{{\rm \min}}}^{1}\frac{d\Gamma_{N\to e^{+}e^{-}\nu_{\tau}}^{{\rm flight}}}{dE_{e}dc_{\theta}}dc_{\theta}\,,\label{eq:-41}
\end{equation}
where $c_{\theta}^{{\rm \min}}$ is given in \cref{eq:-39}. This is followed by the
integration over $E_{N}$ from $E_{N}^{\min}$ to $E_{N}^{\max}$,
given by \cref{eq:-22-1,eq:-23-1}, respectively.

\end{document}